\documentclass[twocolumn,prb,aps,eqsecnum,amsmath,showpacs]{revtex4}
\usepackage{graphicx}
\usepackage{bm}
\usepackage{amsmath}
\begin{document}
\title{Electronic structure and optical properties of quantum-confined lead-salt nanowires}
\author{Valery I. Rupasov}
\email{valery_rupasov@hotmail.com}
\affiliation{ANTEOS, Inc., Shrewsbury, Massachusetts 01545, USA}

\begin{abstract}
In the framework of four-band envelope-function formalism, developed earlier for spherical semiconductor
nanocrystals, we study the electronic structure and optical properties of quantum-confined lead-salt (PbSe
and PbS) nanowires (NWs) with a strong coupling between the conduction and the valence bands. We derive
spatial quantization equations, and calculate numerically energy levels of spatially quantized states of
a transverse electron motion in the plane perpendicular to the NW axis, and electronic subbands developed
due to a free longitudinal motion along the NW axis. Using explicit expressions for eigenfunctions of the
electronic states, we also derive analytical expressions for matrix elements of optical transitions and study
selection rules for interband absorption.

Next we study a two-particle problem with a conventional long-range Coulomb interaction and an interparticle
coupling via medium polarization. We derive analytical expressions for an effective direct Coulomb coupling
and an effective coupling via medium polarization averaging corresponding coupling energies over the fast
transverse motion of charge carriers, and then compute numerically the effective couplings for the lowest-energy
electron-hole pair in a PbSe NW of the radius $R=5$ nm in vacuum. The obtained results show that due to a
large magnitude of the high-frequency dielectric permittivity of PbSe material, and hence, a high dielectric
NW/vacuum contrast, the effective coupling via medium polarization significantly exceeds the effective direct
Coulomb coupling at all interparticle separations along the NW axis.

Furthermore, the strong coupling via medium polarization results in a bound state of the longitudinal motion of
the lowest-energy electron-hole pair (a longitudinal exciton), while fast transverse motions of charge carriers
remain independent of each other. For a PbSe NW of the radius $R=5$ nm, the binding energy of the longitudinal
exciton is found to be about 77.9 meV that approximately two times smaller than the energy of spatial quantization
of the lowest-energy electronic states. Thus, the strong interparticle coupling via medium polarization in
quantum-confined lead-salt NWs significantly modifies the single-particle electronic spectrum, and could result
in essential modifications such Coulomb phenomena as impact ionization, Auger recombination and carrier multiplication.
\end{abstract}


\maketitle

\section{Introduction}

Knowledge of the electronic states is needed for studies of many physical properties such lower-dimensional
quantum-confined semiconductor systems as nanocrystals (NCs) and nanowires (NWs). The electronic states of
NCs have been studied analytically and numerically for both narrow-gap\cite{KW} and wide-gap\cite{ER}
semiconductor materials with different bulk band structures, while the structural, electronic, and optical
properties of quantum-confined NWs have been computed mainly with \emph{ab initio} methods, \cite{RN,BK,DS,RL,MW}
such as density-functional theory, pseudopotentials \cite{YZZ,CZ} and tight-binding methods.\cite{PX1,PX2,PX3,ND}
Theoretical studies of optical properties of NWs made of wide-gap materials, such as GaAs, CdSe, and InP,
have been carried out\cite{M} in the framework of ``particle-in-a-box'' model not accounting for an interband
coupling between the conduction and the valence bands.

In this paper, in the framework of four-band envelope-function formalism, we first obtain analytical expressions
for eigenfunctions of the electronic states of quantum-confined narrow-gap lead-salt (PbSe and PbS) NWs with a strong
coupling between the conduction and the valence bands, and then derive the electronic structure imposing the boundary
condition of vanishing envelope functions on the NW interface. Thus, we generalize an approach developed by Kang and
Wise\cite{KW} for spherical lead-salt NCs to the case of NWs of the cylindrical geometry. However, such a generalization
is not trivial, because it requires new mathematical constructions, specific for the cylindrical geometry, while in
the case of the spherical geometry, conventional mathematical technique developed in the Dirac theory of relativistic
electron \cite{BLP} is employed.

Making use of explicit expressions for electronic eigenfunctions, we derive analytical expressions for
matrix elements of the operator $\bm{e}\cdot\bm{p}$ (where $\bm{e}$ and $\bm{p}$ are the light polarization
and the electron translation-momentum vectors, respectively), which determine optical absorption, and study
selection rules for interband absorption.

Then, we study a long-range Coulomb interaction in NWs. As in the case of NCs, \cite{B1,B2} it contains a
conventional Coulomb interaction, a coupling between a charge particle and a medium polarization created
by this particle itself, and finally an interparticle coupling via medium polarization, i.e. a coupling
between a particle and a medium polarization created by the other particle.

The interaction of a charge particle with its own ``image'' results in a particle repulsion from the NW interface
that effectively reduces the NW radius, and thus slightly modifies the single-particle electronic spectrum.

Since a transverse motion of charge carriers in the plane perpendicular to the NW axis in sufficiently long
NWs of the length $L\gg R$, where $R$ is the NW radius, is obviously much faster than a longitudinal motion
along the NW axis, we derive analytical expressions for an effective direct Coulomb coupling and an effective
coupling via medium polarization averaging the energy of electron-hole ($e$-$h$) coupling with the eigenfunctions
of the transverse motion of charge carriers.

Further numerical calculations for a PbSe NW of the radius $R=5$ nm in vacuum show that owing to a large
magnitude of the high-frequency dielectric permittivity of PbSe material [$\kappa_{\infty}(\text{PbSe}) = 23$],
and hence, high dielectric NW/vacuum contrast, the effective direct Coulomb interparticle coupling is much
weaker then the effective interparticle coupling via medium polarization at all interparticle separations
along the NW axis.

Solving numerically an effective eigenvalue problem for a relative longitudinal motion of the lowest-energy
$e$-$h$ pair, we found a bound state of the longitudinal motion (a longitudinal exciton), and compute its
binding energy, which is found to be about 77.9 meV that is approximately two times smaller than the energy
of spatial quantization of the lowest-energy electronic states. Therefore, in sharp contrast to the case of
NCs, the Coulomb interaction results in essential corrections to the spectroscopy of quantum-confined lead-salt
NWs, and could significantly modify such Coulomb phenomena as impact ionization, Auger recombination and carrier
multiplication.

\section{Four-band envelop function formalism for cylindrical geometry}

In the four-band envelope-function formalism, total electronic wave functions in lead-salt semiconductor materials
are written as a product of the four (for two possible directions of the electron spin in the conduction and
the valence bands) band-edge Bloch functions $|u_i\rangle$ in the conduction ($i=1,2$) and the valence ($i=3,4$)
bands, and four-component envelope functions ${\cal F}_i$,
\begin{equation}
|\psi\rangle = \sum_{i=1}^{i=4}{\cal F}_i|u_i\rangle.
\end{equation}
Boundary conditions in quantum-confined NCs and NWs, are imposed on envelope functions, which are found
as solutions of the eigenvalue problem
\begin{equation}
H{\cal F} = E{\cal F}
\end{equation}
with the eigenenergy $E$ and the Hamiltonian of bulk lead-salt materials in a spherical approximation\cite{KW}
\begin{equation}
H=\left(\begin{array}{cc}\epsilon_c(\bm{p}) &  \eta(\bm{\sigma}\cdot\bm{p})\\
\eta(\bm{\sigma}\cdot\bm{p}) & -\epsilon_v(\bm{p}) \\
\end{array}\right).
\end{equation}
Here $\bm{p}= -i\nabla$ is the wave vector operator applying to the envelop functions,
$\bm{\sigma}=(\sigma_x,\sigma_y,\sigma_z)$,
\begin{equation}
\sigma_x=\left(\begin{array}{cc}
  0 & 1 \\
  1 & 0 \\
\end{array}\right),
\sigma_y=\left(\begin{array}{cc}
  0 & -i \\
  i & 0 \\
\end{array}\right),
\sigma_z=\left(\begin{array}{cc}
  1 & 0 \\
  0 & -1 \\
\end{array}\right)
\end{equation}
are the Pauli matrices, $\eta=\frac{\hbar}{m_0}P$ (where $m_0$ is the free electron mass, and $P$ is
the Kane momentum) is a parameter of the interband coupling,
$$
\epsilon_c(\bm{p}) = \frac{E_{\text{g}}}{2}+\frac{\hbar^2\bm{p}^2}{2m_c},\;\;\;
\epsilon_v(\bm{p}) = \frac{E_{\text{g}}}{2}+\frac{\hbar^2\bm{p}^2}{2m_v}
$$
are the operators of bare (i.e. in the absence of the interband coupling) electron energies in
the conduction ($c$) and the valence ($v$) bands, $E_\text{g}$ is the energy gap, and $m_{c,v}$
are the effective electron masses in the bands.

In the cylindrical coordinates $(r, \phi, z)\equiv(\bm{r},z)$, where the $Z$ axis is directed along
the NW axis, and $\bm{r}$ is the radius-vector in the $XY$ plane perpendicular to the $Z$ axis, it
is convenient to separate the transverse and longitudinal motions and to rewrite the Hamiltonian
as
\begin{subequations}
\begin{equation}
H=H_{xy}+H_z,
\end{equation}
where
\begin{equation}
H_{xy}=\left(\begin{array}{cc}
\epsilon_c(\bm{q}) &  \eta(\bm{\sigma}\cdot\bm{q})\\
\eta(\bm{\sigma}\cdot\bm{q}) & -\epsilon_v(\bm{q}) \\
\end{array}\right)
\end{equation}
and
\begin{equation}
H_z=\left(\begin{array}{cc}
\frac{\hbar^2k_z^2}{2m_c} &  \eta(\sigma_z\cdot k_z)\\
\eta(\sigma_z\cdot k_z) & -\frac{\hbar^2k_z^2}{2m_v} \\
\end{array}\right).
\end{equation}
\end{subequations}
Here, the total wave vector operator $\bm{p}$ is represented as a sum of the wave vector operator
$\bm{q}=-i\frac{\partial}{\partial\bm{r}}$ of the transverse motion and the wave vector operator
$k_z=-i\frac{\partial}{\partial z}$ of the longitudinal motion.

It is convenient to find first solutions of an auxiliary eigenvalue problem with the Hamiltonian
of the transverse motion $H_{xy}$, and to study then the eigenvalue problem (2.2) with the total
Hamiltonian $H$.

\section{Auxiliary eigenvalue problem}

The auxiliary eigenvalue problem
\begin{equation}
H_{xy}\psi = \varepsilon_0\psi
\end{equation}
for a four-component bispinor
$\psi=\left(\begin{array}{c}
  \varphi \\
  \chi \\
\end{array}\right)
$, where $\varphi$ and $\chi$ are two-component spinors, takes the form
\begin{subequations}
\begin{eqnarray}
\left[\varepsilon_0-\epsilon_c(\bm{q})\right]\varphi &=& \eta(\bm{\sigma}\cdot\bm{q})\chi, \\
\left[\varepsilon_0 + \epsilon_c(\bm{q})\right]\chi &=&
\eta(\bm{\sigma}\cdot\bm{q})\varphi.
\end{eqnarray}
\end{subequations}

In the absence of the interband coupling (particle-in-a-box model), $\eta=0$, solutions of
Eqs. (3.2) are easily found to be
\begin{subequations}
\begin{equation}
\varphi,\chi = J_{m_l}(qr)e^{im_l\phi}\xi_\sigma,
\end{equation}
where $J_{m_l}(qr)$ are the Bessel functions,
$$
\xi_+=\left(\begin{array}{c}
  1 \\
  0 \\
\end{array}\right),\;\;
\xi_-=\left(\begin{array}{c}
  0 \\
  1 \\
\end{array}\right)
$$
are the spinors corresponding to two possible projections of the electron spin on the $Z$ axis,
and $m_l=0, \pm 1, \ldots$ are eigenvalues of the orbital angular momentum operator of the
transverse motion
$\bm{l}=\bm{r}\times\bm{q} =-i\left(\bm{r}\times \frac{\partial}{\partial\bm{r}}\right)$ directed
along the $Z$ axis.

Taking into account also a free longitudinal motion with the eigenfunction $e^{ikz}$, the eigenenergies
in the particle-in-a-box model
\begin{equation}
E=\left\{\begin{array}{c}
  +\left(\frac{E_{\text{g}}}{2}+\frac{\hbar^2p^2}{2m_c}\right) \\
  -\left(\frac{E_{\text{g}}}{2}+\frac{\hbar^2p^2}{2m_v}\right) \\
\end{array} \right.
\end{equation}
\end{subequations}
correspond to states in the conduction ($+$) and the valence band ($-$) with the total wave vector
$p=\sqrt{q^2+k^2}$, where $q$ and $k$ are wave vectors of the transverse and the longitudinal motions,
respectively.

The wave functions (3.3a) describe states with given $z$-projections of both spin and orbital angular
momentum of the transverse motion. However, because of the interband spin-orbit coupling term in
the Hamiltonian $H_{xy}$, these projections are not conserved separately, and only the projection
of the total angular momentum $m_j=m_l+s_z$ is conserved.

From functions $e^{im_l\phi}$ and spinors $\xi_{\sigma}$ ($\sigma=\pm$) one can construct two polar
angular spinors:
\begin{subequations}
\begin{equation}
\Omega_{m_j}(\phi)=\frac{i^{m_l}}{\sqrt{2\pi}}e^{im_l\phi}\left(
\begin{array}{c}
  1 \\
  0 \\
\end{array}\right),\;\;\;m_l=m_j-\frac{1}{2},
\end{equation}
and
\begin{equation}
\Theta_{m_j}(\phi)=\frac{i^{m'_l}}{\sqrt{2\pi}}e^{im'_l\phi}\left(
\begin{array}{c}
  0 \\
  1 \\
\end{array}\right),\;\;\;m'_l=m_j+\frac{1}{2},
\end{equation}
\end{subequations}
which are analogous, in some sense, to the angular spinors\cite{BLP} in the spherical geometry.

The polar angular spinors are eigenstates of all three angular operators: the spin operator
$\bm{s}_z=\frac{1}{2}\sigma_z$ with eigenvalues $s_z=\pm\frac{1}{2}$, the total angular momentum
operator $\bm{j}=\bm{l}+\bm{s}_z$ with eigenvalues $m_j$, and the operator $\bm{l}$ with two possible
eigenvalues $m_l=m_j-\frac{1}{2}$ and $m'_l=m_j+\frac{1}{2}$ at a given value of the total angular
momentum projection. Moreover, they are orthonormal,
\begin{subequations}
\begin{eqnarray}
\int_0^{2\pi}d\phi\Omega^\dag_{m_j}(\phi)\Omega_{m'_j}(\phi)&=&\delta_{m_jm'_j},\\
\int_0^{2\pi}d\phi\Theta^\dag_{m_j}(\phi)\Theta_{m'_j}(\phi)&=&\delta_{m_jm'_j},
\end{eqnarray}
\end{subequations}
and orthogonal to each other, $\Theta^\dag_{m'_j}\Omega_{m_j}=0$, owing to orthogonality of the
spinors $\xi_+$ and $\xi_-$.

The introduced polar spinors are related to each other by the expressions
\begin{subequations}
\begin{eqnarray}
\Omega_{m_j}&=&-i(\bm{\sigma}\cdot\bm{e}_r)\Theta_{m_j},\\
\Theta_{m_j}&=&i(\bm{\sigma}\cdot\bm{e}_r)\Omega_{m_j},
\end{eqnarray}
\end{subequations}
where $\bm{e}_r\equiv \bm{r}/r =\cos\phi\cdot\bm{e}_x+\sin\phi\cdot\bm{e}_y$, $\bm{e}_x$ and $\bm{e}_y$
are the unit vectors along the $X$ and $Y$ axis, and hence,
$$
\bm{\sigma}\cdot\bm{e}_r=\cos{\phi}\cdot\sigma_x+\sin{\phi}\cdot\sigma_y=
\left(\begin{array}{cc}
  0 & e^{-i\phi} \\
  e^{i\phi} & 0 \\
\end{array}\right).
$$
The relations (3.6) play an important role in further computations, because they allow one to split
the system of equations (3.2) into independent equations for the bispinor components $\varphi$ and $\chi$.

As in the case of the spherical geometry, \cite{BLP} one can now construct two different bispinors with a
given total angular momentum projection $m_j$ and uncertain values of the orbital angular momentum and
spin projections:
\begin{subequations}
\begin{equation}
\Psi_{m_j} = \left(
\begin{array}{c}
  Af_{m_l}(r)\Omega_{m_j}(\phi) \\
  Bf_{m'_l}(r)\Theta_{m_j}(\phi)\\
\end{array}\right)
\end{equation}
and
\begin{equation}
\Phi_{m_j} = \left(
\begin{array}{c}
  Cf_{m'_l}(r)\Theta_{m_j}(\phi) \\
  Df_{m_l}(r)\Omega_{m_j}(\phi)\\
\end{array}\right),
\end{equation}
\end{subequations}
where the coefficients and radial functions $f_{m_l}$ and $f_{m'_l}$ should be found from solutions
of the eigenvalue problem.

Inserting the expressions (3.7) into Eqs. (3.2) we find after tedious but straightforward computations:
\begin{subequations}
\begin{equation}
\Psi_{+,m_j,q}=\frac{1}{\sqrt{2\varepsilon_0}}\left(
\begin{array}{c}
  \sqrt{\varepsilon_0+\epsilon_q}f_{m_l}(r)\Omega_{m_j}(\phi) \\
  \sqrt{\varepsilon_0-\epsilon_q}f_{m'_l}(r)\Theta_{m_j}(\phi)\\
\end{array}\right),
\end{equation}
\begin{equation}
\Psi_{-,m_j,q}=\frac{1}{\sqrt{2\varepsilon_0}}\left(\begin{array}{c}
  \sqrt{\varepsilon_0-\epsilon_q}f_{m_l}(r)\Omega_{m_j}(\phi) \\
  - \sqrt{\varepsilon_0+\epsilon_q}f_{m'_l}(r)\Theta_{m_j}(\phi)\\
\end{array}\right),
\end{equation}
\end{subequations}
and
\begin{subequations}
\begin{equation}
\Phi_{+,m_j,q}=\frac{1}{\sqrt{2\varepsilon_0}}\left(
\begin{array}{c}
  \sqrt{\varepsilon_0+\epsilon_q}f_{m'_l}(r)\Theta_{m_j}(\phi) \\
  \sqrt{\varepsilon_0-\epsilon_q}f_{m_l}(r)\Omega_{m_j}(\phi)\\
\end{array}\right),
\end{equation}
\begin{equation}
\Phi_{-,m_j,q}=\frac{1}{\sqrt{2\varepsilon_0}}\left(\begin{array}{c}
  \sqrt{\varepsilon_0-\epsilon_q}f_{m'_l}(r)\Theta_{m_j}(\phi) \\
  -\sqrt{\varepsilon_0+\epsilon_q}f_{m_l}(r)\Omega_{m_j}(\phi)\\
\end{array}\right)
\end{equation}
\end{subequations}
in the conduction ($+$) and the valence ($-$) bands, which describe states with
the eigenenergy
\begin{equation}
\varepsilon_0 = +\sqrt{\epsilon_q^2+\eta^2q^2}
\end{equation}
in the conduction band, and $-\varepsilon_0$ in the valence band. To simplify the above
expressions, we used a ``mirror'' symmetry of the conduction and valence bands in PbSe and
PbS materials, and set $m_c=m_v\equiv m$. Therefore, here and hereafter
$\epsilon_q=\frac{E_{\text{g}}}{2}+\frac{\hbar^2q^2}{2m}$.

The radial functions
\begin{equation}
f_{m_l}(r) = \frac{1}{N_{m_l}}\left[J_{m_l}(qr) -
\frac{J_{m_l}(qR)}{I_{m_l}(\lambda R)}I_{m_l}(\lambda r)\right],
\end{equation}
where $I_{m_l}(\lambda r)$ are the modified Bessel functions,
$\lambda=\sqrt{q^2+\lambda_0^2}$, and
$
\lambda_0=\frac{2m}{\hbar^2}\sqrt{\frac{\hbar^2}{2m}E_{\text{g}}
+\eta^2},
$
vanish on the NW interface ($r=R$) at any wave vector $q$. The coefficients $N_{m_l}$
are found from the normalization condition
$$
\int_0^R rdrf^2_{m_l}(r) = 1.
$$
With normalized $f$-functions, the bispinors are obviously normalized to unity.

The condition of self-consistency of simultaneous vanishing both spinor components of the bispinors
on the NW interface results in the spatial quantization equations for the transverse wave vector:
\begin{subequations}
\begin{equation}
\frac{\sqrt{\varepsilon_0+\epsilon_q}}{\sqrt{\epsilon_\lambda-\varepsilon_0}}
\frac{J_{m_l}(qR)}{I_{m_l}(\lambda R)}=
\frac{\sqrt{\varepsilon_0-\epsilon_q}}{\sqrt{\epsilon_\lambda+\varepsilon_0}}
\frac{J_{m'_l}(qR)}{I_{m'_l}(\lambda R)}
\end{equation}
for the bispinors $\Psi_+$ and $\Phi_-$, and
\begin{equation}
\frac{\sqrt{\varepsilon_0+\epsilon_Q}}{\sqrt{\epsilon_\Lambda-\varepsilon_0}}
\frac{J_{m'_l}(QR)}{I_{m'_l}(\lambda R)}=
\frac{\sqrt{\varepsilon_0-\epsilon_Q}}{\sqrt{\epsilon_\Lambda+\varepsilon_0}}
\frac{J_{m_l}(QR)}{I_{m_l}(\lambda R)}
\end{equation}
\end{subequations}
for the bispinors $\Psi_-$ and $\Phi_+$. Here, $\epsilon_\lambda=\epsilon_q+\frac{2m}{\hbar^2}\eta^2$,
$\epsilon_\Lambda=\epsilon_Q+\frac{2m}{\hbar^2}\eta^2$ and we use two notations to emphasize the
existence of two distinct sets of spatially quantized wave vectors, $q_n$ and $Q_n$ ($n=1,2,\ldots$),
corresponding to the same total angular momentum projection $m_j$.

As it must be expected, $\Psi$-bispinors are orthogonal to $\Phi$-bispinors due to orthogonality of
the $\Omega$ and $\Theta$ polar angular spinors, while orthogonality of bispinors corresponding
to states in different bands, $\Psi_+$, $\Psi_-$ and $\Phi_+$, $\Phi_-$, with the same $m_j$ is
provided by the structure of energy factors in the expressions (3.8) and (3.9).

The bispinors describing electronic states in the conduction band contain a contribution of the valence
band, and vice versa. Correspondingly, the total wave functions (2.1) contain contribution of the
band-edge Bloch functions of the conduction and the valence bands. While, in the absence of the interband
coupling, the second components corresponding to the valence band in $\Psi_+$ and $\Phi_+$ bispinors and
the first components corresponding to the conduction band in $\Psi_-$ and $\Phi_-$ bispinors vanish due to
vanishing the factor $\sqrt{\varepsilon_0-\epsilon_q}$ at $\eta\rightarrow 0$.

\section{Longitudinal motion}

The longitudinal motion along the $Z$ axis additionally mixes quantum states in the conduction and the
valence bands. Solutions of the eigenvalue problem with the total Hamiltonian
\begin{equation}
H\psi=(H_{xy}+H_z)\psi=E\psi
\end{equation}
can be found as a linear superpositions of the eigenstates of the transverse motion Hamiltonian $H_{xy}$.

Since the interband coupling term in the Hamiltonian $H_z$ mixes quantum states in the conduction and the
valence bands describing by different bispinors $\Psi$ and $\Phi$, one can look for solutions of the
eigenvalue problem (4.1) in the form
\begin{subequations}
\begin{eqnarray}
F&=& (A\Psi_+ + B\Phi_-)e^{ikz},\\
G&=& (C\Psi_- + D\Phi_+)e^{ikz},
\end{eqnarray}
\end{subequations}
where the coefficients should be found from solution of the eigenvalue problem. Inserting these
expressions into Eq. (4.1) we finally find:
\begin{subequations}
\begin{equation}
F_+ =\frac{1}{\sqrt{2E}}\left( \sqrt{E+\varepsilon}\Psi_+
-\sqrt{E-\varepsilon}\Phi_-\right)e^{ikz}
\end{equation}
\begin{equation}
G_+= \frac{1}{\sqrt{2E}} \left(\sqrt{E-\varepsilon}\Psi_- +
\sqrt{E+\varepsilon}\Phi_+\right)e^{ikz}
\end{equation}
\end{subequations}
for states in the conduction band with the eigenenergy
\begin{equation}
E=+\sqrt{\epsilon_p^2+\eta^2p^2},
\end{equation}
and
\begin{subequations}
\begin{equation}
F_- = \frac{1}{\sqrt{2E}}\left(\sqrt{E-\varepsilon}\Psi_+
+\sqrt{E+\varepsilon}\Phi_-\right)e^{ikz}
\end{equation}
\begin{equation}
G_-=\frac{1}{\sqrt{2E}}\left(\sqrt{E+\varepsilon}\Psi_-
-\sqrt{E-\varepsilon}\Phi_+\right)e^{ikz}
\end{equation}
\end{subequations}
for states in the valence band with the eigenenergy $-E$. Here $\varepsilon=\sqrt{\epsilon_p^2+\eta q^2}$,
$\epsilon_p=\frac{E_{\text{g}}}{2}+\frac{\hbar^2p^2}{2m}$, and $p^2=q^2+k^2$.

It is easy to see that the longitudinal motion does not modify the spatial quantization
equations (3.12), because it mixes bispinors with the same sets of the transverse
wave vectors $q_n$ and $Q_n$.

Thus, eigenstates of the total Hamiltonian $H$ in the cylindrical geometry are characterized
by the projection of the total angular momentum $m_j=\pm\frac{1}{2},\pm\frac{3}{2},\ldots$ on
the $Z$ axis and the continuous wave vector of the longitudinal motion $k$, while the spatially
quantized wave vectors $q_n$ and $Q_n$ of the transverse motion are determined by Eq. (3.12a) for
the bispinors $F_{\pm}$, and by Eq. (3.12b) for the bispinors $G_{\pm}$. The bispinors are
orthogonal to each other and normalized to $2\pi\delta(k-k')$.

Since there are two kinds of bispinors ($F$ and $G$), the number of quantum numbers, which
characterize eigenstates of the Hamiltonian $H$, is equal to that in the absence of the
interband coupling (the particle-in-a-box model), as it must be expected.

Finally, corrections related to mixing of quantum states in the conduction and the valence
bands due to the longitudinal motion are essential only at sufficiently large vectors $k$
comparable to magnitudes of the transverse motion wave vectors $q$ and $Q$. While the densities
of electronic states per unit length of a NW exhibit obviously the Van Hove singularities at
$k\rightarrow 0$:
\begin{equation}
\text{g}_q(E)= \frac{1}{L}\frac{dN}{dE}\sim \frac{1}{2^{3/2}\pi}
\frac{1}{\sqrt{{\frac{\hbar^2}{m}\epsilon_q+\eta^2}}}
\sqrt{\frac{\varepsilon_0(q)}{E-\varepsilon_0(q)}},
\end{equation}
where $N$ is the number of states, and $\varepsilon_0(q)$ are the subband-edge energies determined
in Eq. (3.10).

\begin{figure}[t]
\includegraphics[width=0.57\linewidth]{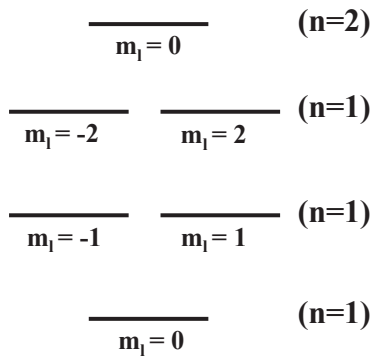}
\caption{Band-edge energy levels ($k=0$) in the conduction band in
the absence of the interband coupling.}
\end{figure}

\section{Electronic structure}

For further comparisons, we consider first the electronic structure in the framework of
particle-in-a-box model, in which the boundary condition $\psi(r=R)=0$ for the wave
functions (3.3a) results in a simple spatial quantization equation
\begin{equation}
J_{m_l}(qR)=0.
\end{equation}
The first three zeroes $j_{m_l,n}$ for $m_l= 0,\pm 1, \pm 2$, where the second index $n=1,2,\ldots$
numerates zeroes of the Bessel functions are approximately found to be:
$$
\begin{array}{ccc}
  j_{0,1}=2.4 & j_{0,2}= 5.5 & j_{0,3}= 8.7 \\
  j_{1,1}=3.8 & j_{1,2}= 7.0 & j_{1,3}= 10.2 \\
  j_{2,1}=5.1 & j_{2,2}= 8.4 & j_{2,3}= 11.6 \\
\end{array}
$$
Then, the subband-edge energies ($k=0$) corresponding to states with the orbital angular momentum projection
$\pm m_l$ are given by:
\begin{equation}
E_{m_l,n} = \frac{E_{\text{g}}}{2}+
\frac{\hbar^2j^2_{m_l,n}}{2mR^2}
\end{equation}
The subband-edge states are additionally degenerate with respect to two possible projection of
the electron spin on the $Z$ axis. Thus, we find for the energies of subbands with the orbital
angular momentum projection $\pm m_l$ and the continuous wave vector $k$
\begin{equation}
E_{\pm,m_l,n}(k) = \pm\left(E_{m_l,n}+\frac{\hbar^2k^2}{2m}\right),
\end{equation}
where the signs $+$ and $-$ correspond to the conduction and the valence band, respectively. The
subband-edge structure in the conduction band is shown in Fig. 1.

The interband coupling completely lifts the degeneration of subband-edge states. The results of
numerical calculations of the electronic structure in the conduction band for a PbSe NW of the 
radius $R = 5$ nm are presented in Table I and illustrated in Fig. 2. For numerical calculations
we adopted the following parameters of PbSe material from Ref. [1]: $E_{\text{g}}=0.28$ eV, 
$m_c=m_v=0.20m_0$, and $\eta = 0.31$ eV$\cdot$nm$^{-1}$.
\begin{figure}[t]
\includegraphics[width=0.77\linewidth]{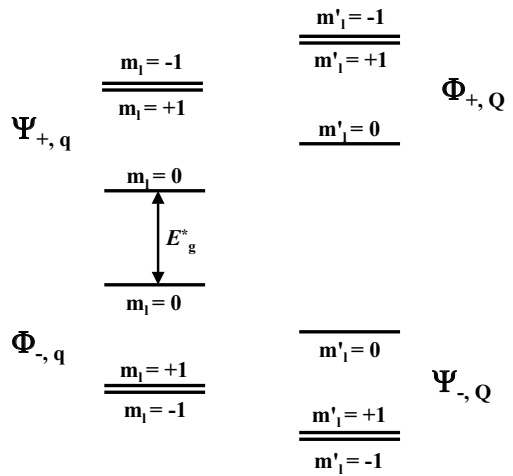}
\caption{Spatially quantized electronic states described by the
bispinors $\Psi_+(q)$ and $\Phi_+(Q)$ in the conduction band, and
mirror symmetric states in the valence band described by the
bispinors $\Phi_-(q)$ and $\Psi_-(Q)$.}
\end{figure}
\begin{table}[h]
\begin{tabular}{lccccc}
\toprule
State ($m_l)$&0&$+1$&-1&+2&-2\\
\colrule
Energy ($\varepsilon_0$)&0.22&0.32&0.33&0.43&0.45\\
\colrule
State ($m'_l$)&0&+1&-1&+2&-2\\
\colrule
Energy ($\varepsilon_0$)&0.25&0.37&0.37&0.49&0.49\\
\botrule
\end{tabular}
\caption{Energies of the spatially quantized states described by bispinors $\Psi_+(q)$
($m_l=0,\pm 1,\pm 2$) and $\Phi_+(Q)$ ($m'_l=0,\pm 1,\pm2$) in the conduction band. The
states $m'_l=\pm 1$ and $m'_l=\pm 2$ are actually not degenerate, but the energy differences
are of the order of a few meV only.}
\end{table}

Although the orbital angular momentum projection is not a good quantum number, it is still
convenient to characterize quantum states by the orbital angular momentum projection of the
first (second) spinor component of bispinors in the conduction (valence) band. The projection
of the total angular momentum $m_j$ is determined by the expressions $m_j=m_l+\frac{1}{2}$
and $m_j=m'_l-\frac{1}{2}$. Thus, the lowest (upper) state in the conduction (valence) band
corresponds to $m_j=\frac{1}{2}$.

Finally, Table II summarizes magnitudes of the effective energy gap in PbSe NWs of different radii
computed in the framework of four-band envelope function formalism (the second row) and in the
particle-in-a-box model, $\eta=0$, (the third row).
\begin{table}[t]
\begin{tabular}{cccccc}
\toprule
$R$\, (nm) &4&5&6&7&8\\
\colrule
$E^*_{\text{g}}$\, (eV) &0.51&0.44&0.40&0.37&0.35\\
\colrule
$E^*_{\text{g}}$\, (eV) ($\eta =0$) &0.42&0.37&0.34&0.32&0.31\\
\botrule
\end{tabular}
\caption{The effective energy gap of PbSe NWs of different radii calculated in the framework of
four-band envelope function formalism (the second row) and particle-in-a-box model, $\eta=0$,
(the third row).}
\end{table}
It is easy to see that the interband coupling results in significant size-dependent corrections
to the electronic structure computed within the framework of the particle-in-a-box model. Note
that the interband coupling decreases the wave vector of the lowest-energy electronic states, but
increases the effective energy gap, while in lead-salt NCs the interband coupling reduces\cite{KW}
$E_\text{g}^*$.

\section{Interband optical transitions}

The strength and selection rules of optical transitions in NWs are determined by the matrix element
\begin{equation}
M = \langle\psi|\bm{e}\cdot\bm{p}|\psi\rangle,
\end{equation}
where $\bm{e}$ is the photon polarization vector and $|\psi\rangle$ are the total electronic
wave function defined in Eq. (2.1).

For the case of interband transitions, the matrix element $M_{cv}$ within a given $L$-valley of
lead-salt material is derived by the method described in Ref. [1] as
\begin{eqnarray}
M_{cv} &=& (\bm{e}\cdot\hat{z})P_l\int d\bm{r}dz{\cal F}^\dag_c(\bm{r},z)
(\sigma_x\otimes\sigma_z){\cal F}_v(\bm{r},z)\nonumber \\
&&+\int d\bm{r}dz{\cal F}^\dag_c(\bm{r},z)(\bm{e}\cdot\bm{p}){\cal F}_v(\bm{r},z),
\end{eqnarray}
where $\otimes$ stands for the direct product. To simplify further expressions, we rewrite Eq.
(6.2) as
$$
M_{cv}=(\bm{e}\cdot\hat{z})P_l\langle{\cal F}_c|\sigma_x\otimes\sigma_z|{\cal F}_v\rangle
+\langle{\cal F}_c|\bm{e}\cdot\bm{p}|{\cal F}_v\rangle
$$
In the first term, $\hat{z}$ stands for one of four equivalent $\langle111\rangle$ directions
in the face-centered cubic lattice, and $P_l$ is the matrix element of the longitudinal Kane
momentum between the band-edge Bloch functions. This term becomes isotropic as result of
summing over all four equivalent $L$ valleys.

Inserting into the first term of Eq. (6.2) the expressions for $F_+$ and $F_-$ bispinors, and
using results of computations of matrix elements in Appendix A, we find
\begin{eqnarray}
M_{F_+F_-}^{(1)}&=&(\bm{e}\cdot\hat{z})P_l\langle F_{+,m'_j,k'}|
\sigma_x\otimes\sigma_z|F_{-,m_j,k}\rangle\nonumber\\
&=& - 2\pi(\bm{e}\cdot\hat{z})P_l\frac{\varepsilon}{E}\delta(k'-k)\delta_{m'_jm_j},
\end{eqnarray}
where the ratio $\varepsilon/E$ can be written as
$$
\frac{\varepsilon}{E}=\sqrt{\frac{\epsilon_p^2+\eta^2q^2}{\epsilon_p^2+\eta^2p^2}}.
$$
This expression is valid only for transitions with the same vectors of the transverse motion $q'_n=q_n$
corresponding to $m'_j=m_j$. Transitions with $q'_n\neq q_n$ are also allowed, but their matrix
elements are smaller due to reducing the radial functions overlap at different $q_n$. The matrix
elements for the ``direct'' transitions from $G_-$ to $G_+$ subbands differ in sign only,
$M_{G_+G_-}^{(1)}=-M_{F_+F_-}^{(1)}$, while the matrix elements $M_{F_+G_-}^{(1)}$ and $M_{G+F_-}^{(1)}$
for allowed ($m'_j=m_j$) ``indirect'' transitions are small at $k\ll q_n$, because they are proportional
to a small factor $\eta k/E_\text{g}^*$.

It should be emphasized that $\delta(k'-k)$ function, which expresses conservation of the $Z$-component
of the wave vector in the system electron plus photon, appears in Eq. (6.3a) because we neglect the small
photon wave vector. If the photon wave vector is taking into account, $\delta$-function takes the form
$\delta(k'-k-K_z)$, where $K_z$ is the projection of the photon wave vector on the $Z$ axis.

Thus, the first term of the matrix element is mainly determined by the longitudinal Kane momentum of bulk
material, while envelope functions determine the ratio $\varepsilon/E$ (for transitions with $q'_n=q_n$),
which equals unity at the subband edges, and is reduced when $k$ grows up. Taking into account Van Hove
singularities in the density of electronic states at $k\rightarrow 0$, we should conclude that the subband-edge
absorption significantly exceeds that at finite $k$.

On the contrary, the polarization-dependent second term of the matrix element in Eq. (6.2) (which is obviously
absent in the particle-in-a-box model) is completely determined by envelope functions. The matrix elements for
the direct transitions $M_{F_+F_-}^{(2)}=\langle F_+|\bm{e}\cdot\bm{p}|F_-\rangle$ and
$M_{G_+G_-}^{(2)}\langle G_+|\bm{e}\cdot\bm{p}|G_-\rangle$ are proportional to $\eta k/E_\text{g}^*$.
Therefore, their contribution to the transition strength is small in comparison with that of the matrix elements
$M_{F_+F_-}^{(1)}$ and $M_{G_+G_-}^{(1)}$. While matrix elements for indirect transitions are found to be at
small $k$:
\begin{subequations}
\begin{eqnarray}
M_{F_+G_-}^{(2)}&=&\langle \Psi_{+,m'_j}|\bm{e}\cdot\bm{q}|\Psi_{-,m_j}\rangle,\\
M_{G_+F_-}^{(2)}&=&\langle \Phi_{+,m'_j}|\bm{e}\cdot\bm{q}|\Phi_{-,m_j}\rangle.
\end{eqnarray}
\end{subequations}
They do not vanish only for transitions with $m'_j=m_j\pm 1$. Finally, matrix elements of the operator
$e_z\cdot k_z$ vanish owing to orthogonality of states in the valence and the conduction bands, and absorption of
light polarized along the NW axis is determined by the matrix elements $M_{F_+F_-}^{(1)}$ and $M_{G_+G_-}^{(1)}$
only.

\section{Coulomb interaction and longitudinal excitons}

Now we are able to consider a two-particle problem in quantum-confined lead-salt NWs. To solve the
two-particle problem is convenient to write the NW Hamiltonian in the absence of an interparticle
coupling in the second-quantization representation as
\begin{equation}
H_0=\sum_q\int\frac{dk}{2\pi} E_q(k)\left[c^\dag_q(k)c_q(k)+h^\dag_q(k)h_q(k)\right].
\end{equation}
Here, the operators $c^\dag_q(k)$ [$c_q(k)$] and $h^\dag_q(k)$ [$h_q(k)$] create (annihilate) an electron
and a hole with a wave vector $k$ in a subband with a spatially quantized wave vector $q$ in the conduction
and the valence band, respectively, and the summation is carried out over all the subbands. In what follows,
we restrict our consideration to the case when both electron and hole belong to the lowest-energy subbands
$F_{+,\frac{1}{2}}$ and $F_{-,\frac{1}{2}}$. That allows one to omit the summation over other subbands in
Eq. (7.1).

For a sufficiently slow longitudinal motion of charge carriers with $k\ll q$, where now $q$ is the wave vector
of the transverse motion in the lowest-energy subbands, the expression for the eigenenergy (4.4) is expended as
\begin{equation}
E_q(k) \simeq \varepsilon_0(q)+\frac{\hbar^2k^2}{2m_z(q)}=\frac{1}{2}E_\text{g}^*+\frac{\hbar^2k^2}{2m_z(q)},
\end{equation}
where
$$
m_z(q)=\frac{1}{2}\frac{E_\text{g}^*}{\epsilon_q+\frac{m}{\hbar^2}\eta^2}m
$$
is the effective mass of the longitudinal motion. Introducing the Fourier transformed charge carrier operators
$$
c_q(z)=\int\frac{dk}{2\pi}c_q(k)e^{ikz};\;\;h_q(z)=\int\frac{dk}{2\pi}h_q(k)e^{ikz},
$$
one can rewrite the free Hamiltonian (7.1) as
\begin{eqnarray}
H_0&=&E_\text{g}^*+\int dz_e c_q^\dag(z_e)\left(-\frac{\hbar^2}{2m_z}\frac{d^2}{dz_e^2}\right)c_q(z_e)\nonumber\\
&&+\int dz_h h_q^\dag(z_h)\left(-\frac{\hbar^2}{2m_z}\frac{d^2}{dz_h^2}\right)h_q(z_h)
\end{eqnarray}

At $k\ll q$, the energy factors $\sqrt{E-\varepsilon}$ in Eqs. (4.3) and (4.5) are also small, and the terms
proportional to $\sqrt{E-\varepsilon}$ can be omitted. Then, the total bispinors $F_{\pm}$ differ from
$\Psi$- and $\Phi$-bispinors in the wave function of a free longitudinal motion only, i.e.
$F_{+}\simeq\Psi_{+}\exp{(ikz)}$ and $F_{-}\simeq\Phi_{-}\exp{(ikz)}$. Correspondingly, the operator of
$e$-$h$ coupling is written as
\begin{subequations}
\begin{equation}
\hat{V}=\int dz_e dz_h c_q^\dag(z_e)c_q(z_e)V(z_e-z_h)h_q^\dag(z_h)h_q(z_h),
\end{equation}
where the effective coupling energy is given by
\begin{eqnarray}
V&=&\int d\bm{r}_ed\bm{r}_h\Psi^\dag_{+,\frac{1}{2}}(\bm{r}_e)\Psi_{+,\frac{1}{2}}(\bm{r}_e)
U(\bm{r}_e,z_e;\bm{r}_h,z_h)\nonumber\\
&&\times\Phi^\dag_{-,\frac{1}{2}}(\bm{r}_h)\Phi_{-,\frac{1}{2}}(\bm{r}_h),
\end{eqnarray}
\end{subequations}
with the coupling energy $U$ determined in Eqs. (B5).

Then, the eigenvalue problem for a relative motion of the lowest-energy $e$-$h$ pair with the total
Hamiltonian $H=H_0+\hat{V}$ reads
\begin{equation}
\frac{\hbar^2}{m_z}\frac{d^2}{d\zeta^2}\psi_{eh}(\zeta)+ \left[\varepsilon_{eh}-V(\zeta)\right]
\psi_{eh}(\zeta)=0,
\end{equation}
where $\zeta=z_e-z_h$, $\varepsilon_{eh}=E_{eh}-E_\text{g}^*$, and $E_{eh}$ is the energy of the pair.
Due to the strong interband coupling, the effective mass of the longitudinal motion, $m_z$, in a PbSe
NW of the radius of 5 nm, is found to be about a half of the effective electron mass $m$, $m_z\simeq 0.51 m$.
\begin{figure}[t]
\includegraphics[width=0.87\linewidth]{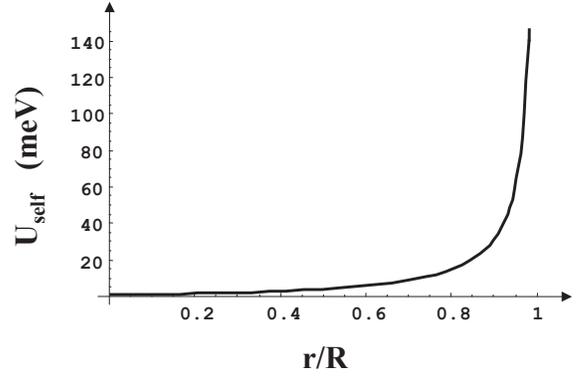}
\caption{The self-interaction energy vs. normalized radial coordinate for a PbSe NW
of $R=5$ nm; the constant term $U_{\text{self}}(r=0)=57.47$ meV is extracted.}
\end{figure}

A charge carrier confined in a cylindrical NW of the dielectric permittivity $\kappa_{\text{nw}}$
placed in a host of the permittivity $\kappa_{\text{h}}$ creates a medium polarization (an
``image'' charge) that results in an interaction between the charge and the image charge. The expression
for their interaction energy, $U_{\text{self}}$, is derived in Appendix B [Eq. (B4)], and the results of
numerical calculations for a PbSe NW of the radius $R=5$ nm in vacuum are plotted in Fig. 3. This
single-particle term must be included in the Hamiltonian (2.3) determining the single-particle electronic
spectrum. Since $U_{\text{self}}(r)$ grows up with $r$ and diverges at $r=R$, the self-interaction repulses
a charge carrier from the interface that obviously results in effective decreasing the NW radius and
increasing (decreasing) electronic energies in the conduction (valence) band. However, the magnitude of
$U_{\text{self}}(r)$ reaches magnitudes of confinement energies only at $r\simeq 0.99R$. Therefore, corrections
to the single-particle spectrum are inessential and can be omitted.

As in the case of NCs, \cite{B1,B2} the energy of total $e$-$h$ interaction $U$ (B5a) is separated
into the energy of direct Coulomb coupling, $U_\text{C}$ [see Eq. (B5b)], and a term corresponding
to the interaction energy between one charge carrier and a medium polarization created by the second
one, $U_\text{pol}$ (B5c). For the lowest-energy $e$-$h$ pair, the direct Coulomb
($V_\text{C}=\langle U_\text{C}\rangle$) and polarization ($V_\text{pol}=\langle U_\text{pol}\rangle$)
parts of the total effective $e$-$h$ coupling are plotted in Fig. 4 as functions of the modulus
of $e$-$h$ separation along the NW axis normalized to the NW radius. It is easy to see that due
to a large magnitude of the dielectric permittivity of PbSe material [$\kappa(\text{PbSe})=23$], and
hence, high dielectric NW/vacuum contrast, the effective coupling via medium polarization $V_\text{pol}$
essentially exceeds the effective direct Coulomb coupling $V_\text{C}$ at all $e$-$h$ separations along
the NW axis. Therefore, the latter can be omitted in the eigenvalue problem (7.5).
\begin{figure}[t]
\includegraphics[width=0.87\linewidth]{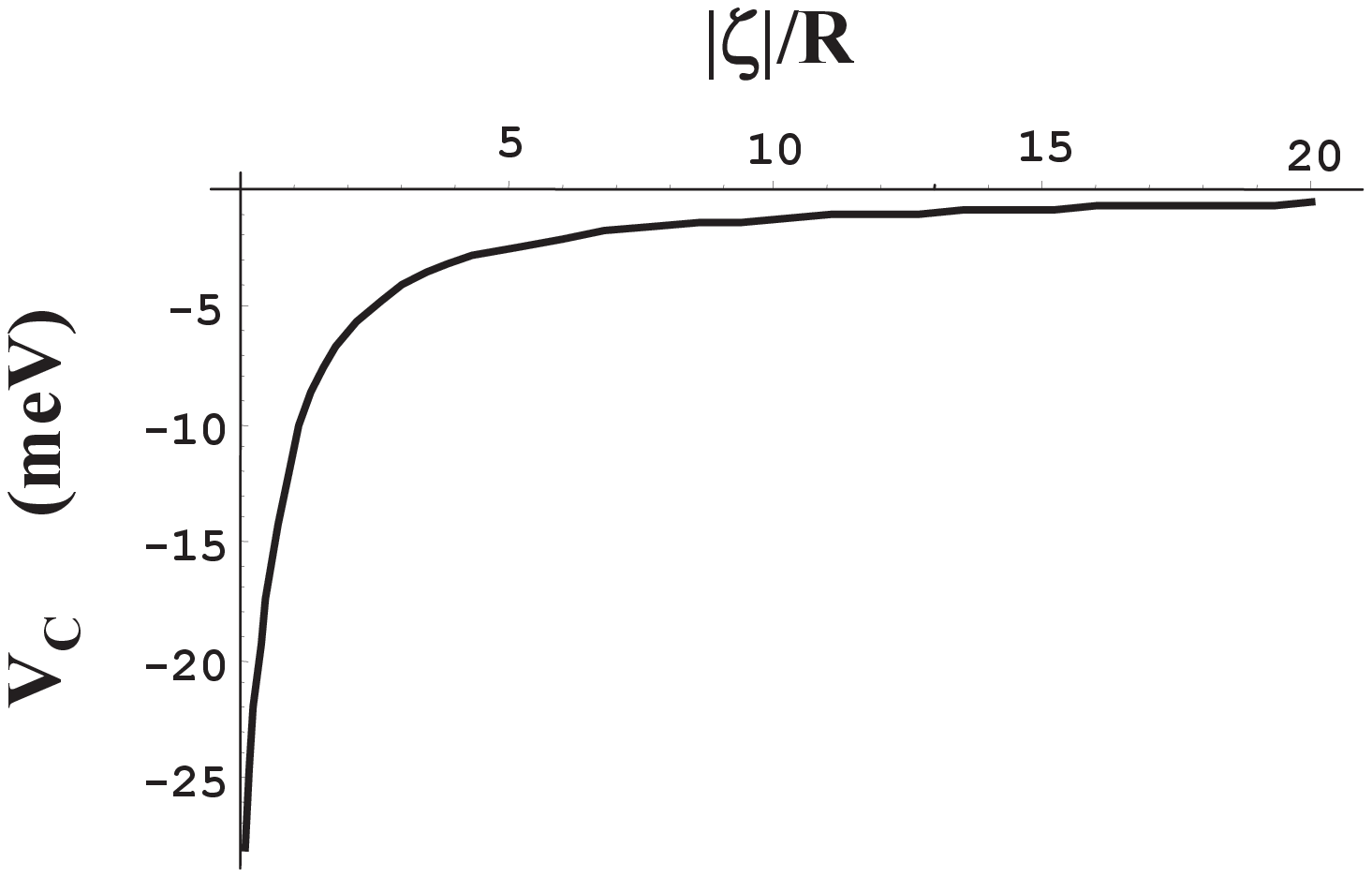}
\includegraphics[width=0.91\linewidth]{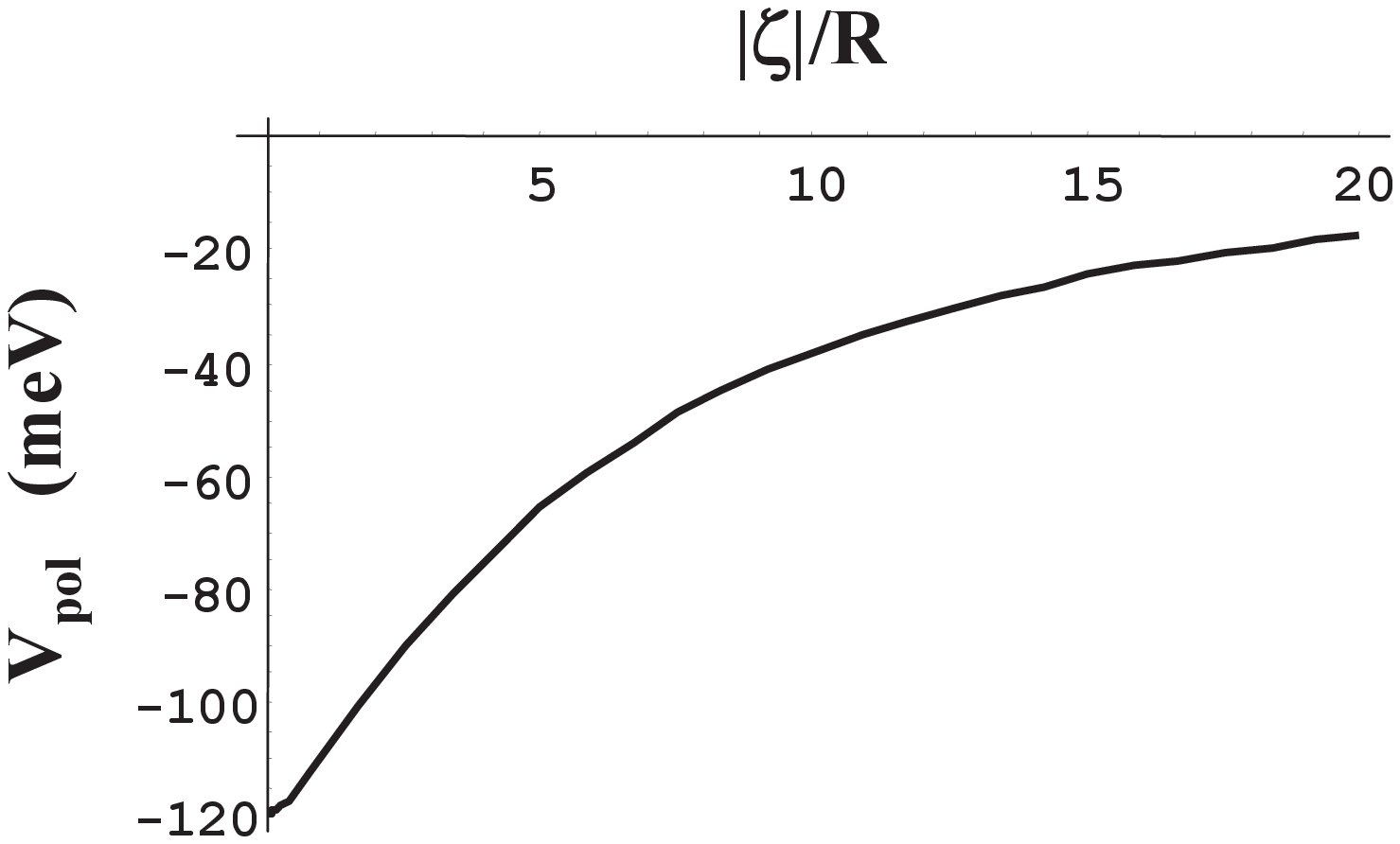}
\caption{Energies of the effective direct Coulomb electron-hole coupling $V_\text{C}$ (top) and the effective
coupling via medium polarization $V_\text{pol}$ (bottom) vs. the normalized modulus of $e$-$h$ separation
$|\zeta|/R=|z_e-z_h|/R$ along the NW axis for a PbSe NW of $R=5$ nm.}
\end{figure}

To avoid too tedious computations with the very complicated function $V_{\text{pol}}(\zeta)$ derived
in Appendix C and plotted in Fig. 4 (bottom), it is convenient to replace $V_{\text{pol}}(\zeta)$
by the function
\begin{equation}
V(\zeta/R)=-\frac{a}{\sqrt{(\zeta/R)^2+b^2}}\;\;\;(\text{meV}),
\end{equation}
which, at $a\approx 392.7$ meV and $b\approx 3.3$, well interpolates the function $V_{\text{pol}}(\zeta)$.
Then, numerical computations show that the lowest-energy electron and hole form a bound state (a longitudinal
exciton) with the exciton binding energy about 77.9 meV, $\varepsilon_{eh}\simeq -77.9$ meV, while fast
transverse motions of charge carriers remain independent of each other. The excitation energy of the
longitudinal exciton, $E_{eh}=E_\text{g}^*+\varepsilon_{eh}$, is then estimated to be about 362 meV,
and hence, it is placed approximately in the middle point between the NW effective energy gap
$E_\text{g}^* = 440$ meV and the energy gap $E_\text{g}=280$ meV of bulk PbSe material.

Finally, the size of the longitudinal exciton is estimated to be about $10R=50$ nm that justifies the
approximation of a slow longitudinal motion in comparison with the transverse motion of charge carriers.
The longitudinal exciton size is comparable to the Bohr radius of Wannier-Mott exciton, $a_B=46$ nm,
in bulk PbSe material. However, such large longitudinal excitons can be destroyed by charge carrier
scattering on impurities and other imperfections in NWs.\cite{N2} Therefore, the problem of longitudinal
excitons and their role in the spectroscopy of quantum-confined lead-salt NWs requires further theoretical
and experimental studies.

\section{conclusions}

In conclusion, in the framework of four-band envelope-function formalism we have studied the electronic structure
and optical properties of quantum-confined lead-salt NWs with a strong coupling between the conduction and the
valence bands. Numerical calculations show that the interband coupling completely lifts the degeneration of electronic
states and results in significant size-dependent corrections to the electronic structure computed in the framework of
particle-in-a-box model. We have also derived analytical expressions for the matrix elements of the operator
$\bm{e}\cdot\bm{p}$, which determine optical absorption in NWs, and have studied selection rules for interband
absorption.

Finally, we have studied a two-particle problem with an effective long-range  Coulomb coupling averaging
corresponding interaction energy over a fast transverse motion of charge carriers. Numerical calculations show that
due to a large magnitude of the dielectric permittivity of lead-salt materials, the effective interparticle coupling
via medium polarization significantly exceeds the effective direct Coulomb coupling.

Furthermore, the strong coupling via medium polarization results in a bound state of the longitudinal motion of the
lowest-energy $e$-$h$ pair (a longitudinal exciton) with the large binding energy comparable to the energy of spatial
quantization of electronic states, and the size about 50 nm, which is comparable to the Bohr radius of Wannier-Mott
exciton in bulk PbSe material. Thus, the strong Coulomb coupling in lead-salt NWs results in significant two- and
many-particle corrections to the single-particle electronic spectrum, and could essentially modify such Coulomb
phenomena as impact ionization, Auger recombination and carrier multiplication.

\acknowledgments

I would like to thank Victor Klimov, who has initiated these theoretical studies, and V. M. Agranovich for numerous
useful discussions and remarks.

\appendix

\section{Matrix elements}

Here, we compute the matrix elements of the operators $\sigma_x\otimes \sigma_z$ and $\bm{e}\cdot\bm{p}$
on the $\Psi$- and $\Phi$-bispinors, which are needed to compute the first and the second terms in the
expression (6.2), and to derive the selection rules for optical transitions.

Note that owing to orthogonality of the $\Omega$- and $\Theta$-spinors and the structure of the operator
$\sigma_x\otimes \sigma_z$, ``diagonal'' matrix elements vanish, i.e.
\begin{equation}
\Psi^\dagger(\sigma_x\otimes \sigma_z)\Psi=\Phi^\dagger(\sigma_x\otimes \sigma_z)\Phi=0
\end{equation}
at any quantum numbers of the bispinors in both the conduction and the valence band. While ``non-diagonal''
matrix elements $\langle\Psi_+|\sigma_x\otimes\sigma_z|\Phi_-\rangle $ and $\langle\Phi_+|\sigma_x\otimes\sigma_z|\Psi_-\rangle$
are found to be:
\begin{eqnarray}
\langle\Psi_{+,m'_j}|\sigma_x\otimes \sigma_z|\Phi_{-,m_j}\rangle&=& -\delta_{m'_j,m_j}, \\
\langle\Phi_{+,m'_j}|\sigma_x\otimes \sigma_z|\Psi_{-,m_j}\rangle&=& \delta_{m'_j,m_j},
\end{eqnarray}
where we used Eqs. (3.5). Therefore, the matrix elements of the operator
$\sigma_x\otimes \sigma_z$ are diagonal in the quantum number $m_j$ and correspond obviously
to the direct interband transitions. Note that these simple expressions are derived for transitions
between states with the same wave vectors $q_n'=q_n$.

On the contrary, owing to orthogonality of polar angular spinors $\Omega$ and $\Theta$, non-diagonal matrix
elements $\langle\Psi_{+,m'_j}|\bm{e}\cdot\bm{p}|\Phi_{-,m_j}\rangle$ and
$\langle\Phi_{+,m'_j}|\bm{e}\cdot\bm{p}|\Psi_{-,m_j}\rangle$ vanish at any quantum numbers,
\begin{equation}
\Psi^\dagger(\bm{e}\cdot\bm{p})\Phi=\Phi^\dagger(\bm{e}\cdot\bm{p})\Psi=0.
\end{equation}
While diagonal matrix elements $\langle\Psi_{+,m'_j}|\bm{e}\cdot\bm{p}|\Psi_{-,m_j}\rangle$ and
$\langle\Phi_{+,m'_j}|\bm{e}\cdot\bm{p}|\Phi_{-,m_j}\rangle$ do not vanish, and contribute to the
strength of optical transitions.

\section{Coulomb interaction}

To our best knowledge, an expression for the Coulomb interaction in NWs was derived for the first time
in Ref. [14]. However, the derived expression contains some mathematical incorrectness, so that electrical
potentials are complex rather than real as it must be. Therefore, here we present a correct solution of
the Coulomb problem, and derive an asymptotic expression for particle-particle coupling at interparticle
separations along the $Z$ axis, essentially exceeding the NW radius, $|z-z'|\gg R$.

An electrical potential $P(\bm{r},z)$ at the point with the coordinates $(\bm{r},z)$ created by an electrical
charge $e$ positioned at the point $(\bm{r}',z')$ inside a NW of permittivity $\kappa_\text{nw}$ placed
in a host medium of permittivity $\kappa_\text{h}$ is determined by the Poisson equation
\begin{subequations}
\begin{equation}
\kappa_{\text{nw}}\triangle P(\bm{r},z)= - 4\pi
e\delta(\textbf{r}-\textbf{r}')\delta(z-z')
\end{equation}
inside the NW ($r<R$), and by the Laplace equation
\begin{equation}
\triangle P(\bm{r},z)= 0
\end{equation}
\end{subequations}
outside the NW ($r>R$), where $\triangle$ is the Laplace operator. Solving these equations with the conventional
boundary conditions \cite{LL}
\begin{subequations}
\begin{eqnarray}
P(r=R-0)&=&P(r=R+0), \\
\kappa_{\text{nw}}\frac{d}{dr}P(r=R-0)&=&\kappa_{\text{h}}\frac{d}{dr}P(r=R+0),
\end{eqnarray}
\end{subequations}
we find for the potential inside the NW ($r<R$)
\begin{subequations}
\begin{equation}
P_{\text{in}}(\bm{r},z)=P_{\text{C}}(\bm{r},z)+P_{\text{pol}}(\bm{r},z),
\end{equation}
where
\begin{eqnarray}
P_{\text{C}}(\bm{r},z)&=& \frac{e}{\kappa_{\text{nw}}}
\frac{1}{\sqrt{|\bm{r}-\bm{r}'|^2+(z-z')^2}}\nonumber\\
&=&\frac{4e}{\kappa_{\text{nw}}}
\sum_{m=-\infty}^{\infty}e^{im(\phi-\phi')}\int_0^{\infty}\frac{dk}{2\pi}
\cos{[k(z-z')]}\nonumber\\
&&\times\left\{
\begin{array}{cc}
I_m(kr)K_m(kr'), & r<r' \\
K_m(kr)I_m(kr'), & r>r'
\end{array}
\right.
\end{eqnarray}
is the Coulomb potential, and
\begin{eqnarray}
P_{\text{pol}}(\bm{r},z) & = & \frac{4e}{\kappa_{\text{nw}}}
\left(1-\frac{\kappa_{\text{h}}}{\kappa_{\text{nw}}}\right)
\sum_{m=-\infty}^{\infty}e^{im(\phi-\phi')}\nonumber\\
&&\times\int_0^{\infty}\frac{dk}{2\pi}\cos{[k(z-z')]}
Q_m(kR)\nonumber\\
&&\times I_m(kr)I_m(kr')
\end{eqnarray}
\end{subequations}
is the potential created by the medium polarization, or, in other words, the potential of an
``image'' charge. Here
$$
Q_m(kR)=-\frac{K_m'(kR)K_m(kR)}{K_m(kR)I'_m(kR)-
\frac{\kappa_{\text{h}}}{\kappa_{\text{nw}}}K'_m(kR)I_m(kR)},
$$
$I_m$ and $K_m$ are the modified Bessel functions, and $I'_m(z)=\frac{d}{dz}I_m(z)$, $K'_m(z)=\frac{d}{dz}K_m(z)$.

Note that integration over $k$ in the Fourier integrals in Eqs. (B3), and hence, in all further expressions
of this kind, is performed over positive values of $k$ only, that eliminates imaginary values related
to the imaginary part of the functions $K_m$ at negative argument.

As in the case of nanocrystals,\cite{B1,B2} the interaction energy of two particles with the charges $e$
and $e'$ in a NW is separated into a self-interaction energy, i.e. the energy of interaction of a charge
with its own image:
\begin{eqnarray}
U_\text{self}(r)&=&\frac{1}{2}eP_{\text{pol}}(\bm{r}=\bm{r}',z=z')
=\frac{2e^2}{\kappa_{\text{nw}}}
\left(1-\frac{\kappa_{\text{h}}}{\kappa_{\text{nw}}}\right)\nonumber\\
&&\times\sum_{m=-\infty}^{\infty}\int_0^{\infty}\frac{dk}{2\pi}
Q_m(kR)I^2_m(kr),
\end{eqnarray}
and the energy of interparticle coupling
\begin{subequations}
\begin{equation}
U=U_\text{C}+ U_\text{pol}.
\end{equation}
Here, the first term is the direct Coulomb coupling energy
\begin{eqnarray}
U_{\text{C}}&=& \frac{ee'}{\kappa_{\text{nw}}}
\frac{1}{\sqrt{|\bm{r}-\bm{r}'|^2+(z-z')^2}}\nonumber\\
&=&\frac{4ee'}{\kappa_{\text{nw}}}
\sum_{m=-\infty}^{\infty}e^{im(\phi-\phi')}\int_0^{\infty}\frac{dk}{2\pi}
\cos{[k(z-z')]}\nonumber\\
&&\times\left\{
\begin{array}{cc}
I_m(kr)K_m(kr'), & r<r' \\
K_m(kr)I_m(kr'), & r>r'
\end{array}
\right.,
\end{eqnarray}
while the second one is the energy of interparticle coupling via a medium polarization, i.e. an interaction
of one charge with the medium polarization created by the second one
\begin{eqnarray}
\!\!\!\!\!U_{\text{pol}}& = & \frac{4ee'}{\kappa_{\text{nw}}}
\left(1-\frac{\kappa_{\text{h}}}{\kappa_{\text{nw}}}\right)
\sum_{m=-\infty}^{\infty}e^{im(\phi-\phi')}\int_0^{\infty}\frac{dk}{2\pi}\nonumber\\
&&\times\cos{[k(z-z')]}Q_m(kR)I_m(kr)I_m(kr').
\end{eqnarray}
\end{subequations}

At large interparticle separations $|z-z'|\gg R$, the leading asymptotic term of the polarization part of
the interparticle coupling is found to be
\begin{equation}
U_{\text{pol}} \sim \frac{ee'}{\kappa_{\text{eff}}}
\frac{1}{|z-z'|},
\end{equation}
where the effective permittivity is given by
$$
\frac{1}{\kappa_{\text{eff}}}=
\frac{1}{\kappa_{\text{h}}}-\frac{1}{\kappa_{\text{nw}}}.
$$
While the asymptotic of the direct Coulomb coupling reads
\begin{equation}
U_{\text{C}} \sim \frac{ee'}{\kappa_{\text{nw}}} \frac{1}{|z-z'|}.
\end{equation}

\section{Effective electron-hole coupling}

Inserting into Eq. (7.4b) the expression for the energy of direct electron-hole Coulomb coupling (B5b),
we find for the effective Coulomb coupling
\begin{subequations}
\begin{eqnarray}
V_\text{C}(\tau) &=&-\frac{4e^2}{\kappa_{\text{nw}}R}\int_0^{\infty}\frac{dk}{2\pi}\cos{(k\tau)}
\int_0^1x_1x_2dx_1dx_2\nonumber\\
&&\times D(x_1)K_0(kx_1)I_0(kx_2)D(x_2),
\end{eqnarray}
where
$$
D(x)=\frac{1}{2\varepsilon_0}\left[(\varepsilon_0+\epsilon_q)f^2_0(x)
+(\varepsilon_0-\epsilon_q)f^2_1(x)\right],
$$
and we introduced dimensionless variables $x_1=r_e/R$, $x_2=r_h/R$, and $\tau=|z_e-z_h|/R$.
Note that the angle integration in Eq. (7.4b) selects the only nonzero term with $m=0$ from the sum
over $m$ in Eq. (B5b). Integrating\cite{GR} over $k$ in Eq. (C1a), we finally derive
\begin{eqnarray}
V_\text{C}(\tau) &=&-\frac{e^2}{\kappa_{\text{nw}}R}
\int_0^1x_1dx_1x_2dx_2 D(x_1)D(x_2)\nonumber\\
&&\times\frac{_2F_1\left(\frac{3}{4},\frac{1}{4};1;\frac{4x_1^2x_2^2}{(x_1^2+x_2^2+\tau^2)^2}\right)}
{\sqrt{x_1^2+x_2^2+\tau^2}},
\end{eqnarray}
\end{subequations}
where $_2F_1$ is the hypergeometric function.

Analogous computations for the polarization part of the effective $e$-$h$ coupling result in
\begin{eqnarray}
V_{\text{pol}}(\tau) & = & -\frac{4e^2}{\kappa_{\text{nw}}R}\left(1-\frac{\kappa_{\text{h}}}{\kappa_{\text{nw}}}\right)
\int_0^\infty\frac{dk}{2\pi}\cos(k\tau)\nonumber\\
&&\times Q_0(k)J^2(k),
\end{eqnarray}
where $J(k)=\int_0^1xdx D(x)I_0(kx)$.

\end{document}